\documentclass[12pt]{article}
\usepackage{amsmath}
\usepackage{cite}

\csname @addtoreset\endcsname{equation}{section}
\newcommand{\eq}{\begin{equation}}
\newcommand{\feq}{\end{equation}}
\newcommand{\eqn}{\begin{eqnarray}}
\newcommand{\feqn}{\end{eqnarray}}
\newcommand{\arr}{\begin{eqnarray*}}
\newcommand{\farr}{\end{eqnarray*}}

\newcommand{\m}{{(m)}}

\newcommand{\ft}[2]{{\textstyle\frac{#1}{#2}}}
\input{tcilatex}

\begin{document}

\begin{titlepage}
\begin{flushright}
CAMS/03-06\\
MCTP-03-37\\
hep-th/0307300
\end{flushright}

\vspace{.3cm}
\begin{center}

\renewcommand{\thefootnote}{\fnsymbol{footnote}}
{\Large \bf Charged configurations in (A)dS spaces}\\

\vspace{1cm}

{\large \bf {James T. Liu$^1$\footnote{email: jimliu@umich.edu} and
W.~A.~Sabra$^2$\footnote{email: ws00@aub.edu.lb}}}\\

\renewcommand{\thefootnote}{\arabic{footnote}}
\setcounter{footnote}{0}

\vspace{1cm}

{\small
$^1$ {Randall Laboratory, Department of Physics, University of Michigan,\\
Ann Arbor, MI 48109--1120}\\
\vspace*{0.4cm}
$^2$ Center for Advanced Mathematical Sciences (CAMS)\\
and\\
Physics Department, American University of Beirut, Lebanon.}
\end{center}

\vspace{1.5cm}
\begin{center}
{\bf Abstract}
\end{center}

We construct new backgrounds of $d$-dimensional gravity with a negative
cosmological constant coupled to a $m$-form field strength.  We find a
class of magnetically charged anti-de Sitter black holes with
$m$-dimensional Einstein horizon of positive, zero or negative curvature.
We also construct a new magnetic co-dimension four brane for the
case of $m=3$.  This brane obeys a charge quantization condition of the
form $q\sim L^2$ where $q$ is the magnetic $3$-form charge and $L$ is
the AdS radius.  In addition, we work out some time-dependent solutions.

\end{titlepage}


\section{Introduction}

Many recent developments in M-theory have resulted from the AdS/CFT
conjecture and its generalizations. In such models, the (at least
asymptotically) anti-de Sitter geometry appears to play an important
r\^{o}le in the proper formulation of a gauge/gravity dual (see,
\textit{e.g.}, \cite{Aharony: 1999ti} and references therein). Hence there
has been much interest in obtaining a better understanding of supergravity
backgrounds that are asymptotic to AdS. An important class of such solutions
include Schwarzschild-AdS as well as $R$-charged black holes. In general,
such black holes, especially the non-extremal solutions, have dual
descriptions corresponding to gauge theories at finite temperature. This
has led to new insights on the thermodynamics of gauge theories, as well
as signatures of phase transitions.

In addition to AdS black holes \cite
{blackholes,bha,bhb,bhc,Liu:1999ai,bhd,bhe,bhf,bhg,bhh,bhi,bhj}, backgrounds
which interpolate between one or more extrema of gauged supergravity have
also received attention \cite{Behrndt,bea,beb,bec,bed,bee,bef,beg,beh,cks}.
For such solutions, one may isolate a `radial' direction and write the bulk
metric in a warped product form
\begin{equation}
ds^{2}=e^{2A(r)}g_{\mu \nu }dx^{\mu }dx^{\nu }+dr^{2}.
\label{eq:warpprod}
\end{equation}
On the gauge theory side, such solutions may be given a physical
interpretation in terms of a renormalization group flow, where $r$ may be
interpreted as an energy scale \cite{deBoer:1999xf}. Of course, this class
of solutions is not entirely distinct from that of AdS black holes; both
cases may be written in terms of a radial $r$-coordinate, and in general the
near-horizon behavior of a black hole (or brane) reduces to the product of a
metric of the form (\ref{eq:warpprod}) times an `internal' space of
positive, zero or negative curvature.

In general, we are interested in developing a systematic treatment of brane
solutions in AdS. However, with the exception of black holes and domain
walls, little is known about such objects. One of the reasons for this is
that, unlike for solutions which are asymptotically Minkowski, the AdS
background introduces a second length scale (namely the AdS radius) in
addition to the scale of the object itself (\textit{e.g.} the Schwarzschild
radius). This results in more complicated profiles for the explicit
solutions. In addition, extended objects in AdS would appear to be sensitive
to the cosmological force originating from the curvature of spacetime. This
creates some difficulties with supersymmetry, although ones that may be
overcome, since special classes of strings and branes in AdS have been
constructed \cite{cks,strings,stringsa,stringsb} (see also 
\cite{Acharya:2000mu,Gauntlett:2000ng,Gauntlett:2001jj,Gauntlett:2001qs}).

Motivated by this desire to more fully develop backgrounds of relevance to
AdS/CFT, in this paper, we examine a large class of magnetic solutions in
anti-de Sitter space. One important note, however, is that while gauged
supergravity is perhaps most relevant for AdS/CFT, for the most part we
forgo supersymmetry altogether. It is of course known that magnetically
charged black holes in AdS are generically non-supersymmetric. Hence it
would be reasonable to expect that we must similarly give up supersymmetry
when considering more general magnetic branes. To motivate our setup, we
recall that in the asymptotically Minkowski case, the relevant fields
supporting a $p$-brane solution are generally a dilatonic scalar $\phi $ and
$m$-form field strength $F_{(m)}$ (carrying either electric or magnetic
charge) as well as the metric $g_{\mu \nu }$. The system may be described by
an effective Lagrangian of the form
\begin{equation}
e^{-1}\mathcal{L}=R-\ft12(\partial\phi)^2
-{\frac{1}{2\cdot m!}}e^{a\phi }F_{(m)}^{2},
\label{eq:amodel}
\end{equation}
where $a$ is a constant parametrizing the dilaton coupling. While this is
not necessarily a complete model in itself, these fields may be suitably
embedded in an appropriate supergravity framework, provided $a$ is chosen
accordingly. In this manner, the resulting $p$-branes may be seen as
supersymmetric objects preserving some fraction of supersymmetry.

In the present case, to obtain solutions which are asymptotically AdS, we
simply add a negative cosmological constant to the system (\ref{eq:amodel}).
Although in general the dilaton cannot be completely ignored, we
nevertheless choose as a simplification to turn off the dilaton. Even in
this case, we find a rich set of solutions, which presumably capture the
main features of the magnetic brane solutions.

The first class of solutions we obtain are $d$-dimensional magnetically
charged AdS black holes that are simply the magnetic duals of the well known
electrically charged Reissner-Nordstrom-AdS black holes. In some cases,
these black holes may be viewed as magnetic solutions in a gauged
supergravity context \cite{romans}. However, except for the charged
quantized black holes which we indicate, the magnetic solutions generally
break all supersymmetries.

We also obtain a more interesting class of magnetic co-dimension four branes
supported by a $3$-form field strength. These solutions satisfy a charge
quantization relation of the form $q\sim L^2$ where $q$ is the magnetic
charge and $L$ is the AdS radius. This is similar to the case of magnetic
co-dimension three branes found in \cite{sabra2,sabra2a}.

In addition to the static brane solutions, it is also straightforward to
construct new cosmological backgrounds by taking an appropriate
time-dependent ansatz. We are interested in such spaces where the $m$-form
field strength has a magnetic flux turned on. In general, these backgrounds
may be obtained by analytic continuation of their static counterparts. We
construct new cosmological solutions in de Sitter space as well as in
anti-de Sitter space.

In the following section, we outline the basic setup for obtaining magnetic
$m$-form solutions, and explicitly construct $d$-dimensional magnetically
charged AdS black holes. In section three, we generalize this by considering
branes with extended longitudinal dimensions. In particular, we construct
the above mentioned class of co-dimension four branes. In section four, we
examine possible near horizon brane geometries. Then in section five we turn
to the case of time-dependent backgrounds. Finally, in the last section, we
conclude with some speculation on the embedding of these solutions in gauged
supergravities, and the possibility of thereby obtaining new supersymmetric
backgrounds with flux.


\section{Black holes with magnetic charge}

Our starting point is $d$-dimensional gravity with a $m$-form field
strength, $F_{(m)}=dA_{(m-1)}$, and negative cosmological constant. The
Lagrangian has the form
\begin{equation}
e^{-1}\mathcal{L}=R-{\frac{1}{2\cdot m!}}F_{(m)}^{2}+(d-1)(d-2)g^{2},
\label{eq:mlag}
\end{equation}
with resulting equations of motion
\begin{eqnarray}
R_{MN} &=&S_{MN}\equiv {\frac{1}{2(m-1)!}}\left( F_{(m)\,MN}^{2}-{\frac{m-1}
{m(d-2)}}g_{MN}F_{(m)}^{2}\right) -(d-1)g^{2}g_{MN},  \notag \\
d\ast F_{(m)} &=&0.
\end{eqnarray}
For $m=2$, this reduces to the familiar Einstein-Maxwell system with a
cosmological constant. In this case, (\ref{eq:mlag}) may often be
interpreted in the context of $d$-dimensional gauged supergravity. However,
we are mainly interested in higher form field strengths, where there is no
obvious supergravity generalization. Because of the absence of
supersymmetry, we are forced to work with the second order equations of
motion when looking for solutions to this model.

We begin with a construction of magnetic black holes in this system. These
may be viewed as charged generalizations of the models found in \cite
{birmingham}. Since we seek a magnetic configuration with $m$-form field
strength, we take a natural metric ansatz
\begin{equation}
ds^{2}=-e^{2A(r)}dt^{2}+e^{2B(r)}dr^{2}+r^{2}h_{ij}(y)dy^{i}dy^{j},
\label{eq:bhmetans}
\end{equation}
where $i,j=1,\ldots,m$ are $m$ directions on an Einstein manifold
$\mathcal{M}^m$, with metric $h_{ij}$ depending on the coordinates $y^{i}$.
In particular, we take
\begin{equation}
R_{ij}(h)=k(m-1)h_{ij},
\label{eq:meins}
\end{equation}
where $k=1,0,-1$ corresponds to elliptic, flat and hyperbolic horizon
metrics. Clearly the dimension of our spacetime is given by $d=m+2$.

For the metric ansatz (\ref{eq:bhmetans}), the non-vanishing Ricci
components are
\begin{eqnarray}
R_{tt} &=&e^{2A-2B}\left( A^{\prime \prime }+A^{\prime\,2}-A^{\prime
}B^{\prime }+\frac{m}{r}A^{\prime }\right) ,  \notag \\
R_{rr} &=&-A^{\prime \prime }-A^{\prime \,2}+A^{\prime }B^{\prime }+
{\frac{m}{r}}B^{\prime },  \notag \\
R_{ij} &=&R_{ij}(h)+e^{-2B}h_{ij}\left( {r}B^{\prime }-rA^{\prime }-m+1
\right),
\label{eq:n1eins}
\end{eqnarray}
where $R_{ij}(h)$ is given in (\ref{eq:meins}). For a magnetic solution, we
take
\begin{equation}
F_{(m)}=q\,d^{m}y,
\label{mf}
\end{equation}
where $d^{m}y$ is the volume form on $\mathcal{M}^{m}$. This yields
\begin{equation}
F_{(m)\,ij}^{2}=(m-1)!\frac{q^{2}}{r^{2(m-1)}}h_{ij},\qquad F_{(m)}^{2}=m!
\frac{q^{2}}{r^{2m}}.
\label{eq:fsqs}
\end{equation}
For this black hole ansatz, we therefore have
\begin{eqnarray}
S_{tt} &=&e^{2A}\left( {\frac{\left( d-3\right) q^{2}}{2(d-2)r^{2(d-2)}}}
+(d-1)g^{2}\right) ,  \notag \\
S_{rr} &=&-e^{2B}\left( {\frac{\left( d-3\right) q^{2}}{2(d-2)r^{2(d-2)}}}
+(d-1)g^{2}\right) ,  \notag \\
S_{ij} &=&-r^{2}\,h_{ij}\left( -{\frac{q^{2}}{2(d-2)r^{2(d-2)}}}
+(d-1)g^{2}\right) ,
\end{eqnarray}
where we have made use of the fact that $d=m+2$. Since the magnetic ansatz
automatically solves the $F_{(m)}$ equation of motion, we only need to worry
about the Einstein equations. These equations are
\begin{eqnarray}
A^{\prime \prime }+A^{\prime \,2}-A^{\prime }B^{\prime }+\frac{d-2}{r}
A^{\prime } &=&e^{2B}\left( {\frac{\left( d-3\right) q^{2}}{2(d-2)r^{2(d-2)}}}
+(d-1)g^{2}\right) ,  \notag
\label{eq:eom} \\
A^{\prime \prime }+A^{\prime\,2}-A^{\prime }B^{\prime }-\frac{d-2}{r}
B^{\prime } &=&e^{2B}\left( {\frac{\left( d-3\right) q^{2}}{2(d-2)r^{2(d-2)}}}
+(d-1)g^{2}\right) , \\
\frac{{(}A^{\prime }-B^{\prime })}{r}+\frac{{d-3}}{r^{2}} &=&e^{2B}\left(
-{\frac{q^{2}}{2(d-2)r^{2(d-2)}}}+(d-1)g^{2}+\frac{k(d-3)}{r^{2}}\right) .
\notag
\end{eqnarray}

In order to solve the Einstein equations, we note that the first two
equations in (\ref{eq:eom}) imply the simple condition
\begin{equation}
A^{\prime }+B^{\prime }=0.
\end{equation}
Combining this with the last equation of (\ref{eq:eom}), we obtain the
following expression
\begin{equation}
\partial _{r}\left( r^{d-3}e^{-2B}\right) =-\frac{q^{2}}{2(d-2)}\frac{1}
{r^{(d-2)}}+(d-1)g^{2}r^{(d-2)}+k(d-3)r^{d-4},
\end{equation}
which can be easily integrated to give the solution
\begin{eqnarray}
e^{-2B} &=&k-\frac{\mu _{0}}{r^{d-3}}+\frac{q^{2}}{2(d-2)(d-3)}\frac{1}
{r^{2(d-3)}}+g^{2}r^{2},  \notag \\
e^{2A} &=&e^{2A_{0}}e^{-2B}.
\end{eqnarray}
While the general solution has two integration constants, $A_0$ and $\mu_0$,
the former may be set to zero by a simple rescaling of the time coordinate.
Thus we are left with a single non-extremality parameter $\mu_0$, as well as
the magnetic charge $q$. The magnetic black hole solution then has the
standard form
\begin{eqnarray}
ds^{2} &=&-f(r)dt^{2}+{\frac{dr^{2}}{f(r)}}+r^{2}h_{ij}(y)dy^{i}dy^{j},
\notag \\
F_{(d-2)} &=&q\,d^{d-2}y,
\end{eqnarray}
where
\begin{equation}
f(r)=k-\frac{\omega _{d}M}{r^{d-3}}+\frac{q^{2}}{2(d-2)(d-3)}\frac{1}
{r^{2(d-3)}}+g^{2}r^{2}.
\label{eq:ffunc}
\end{equation}
Following \cite{birmingham}, we have defined
\begin{equation}
\omega _{d}={\frac{16\pi G}{(d-2)\mathrm{Vol}(\mathcal{M}^{d-2})}},
\end{equation}
where Newton's constant has been restored. For vanishing charge, $q=0$, this
reduces to the solution of \cite{birmingham}.

It ought to be apparent that these black holes are the magnetic duals of the
well-known electrically charged solutions of the Einstein-Maxwell system
with cosmological constant, given by the Lagrangian
\begin{equation}
e^{-1}\mathcal{L}=R-{\frac{1}{4}}F_{(2)}^2+(d-1)(d-2)g^2.
\end{equation}
The electric charged black holes are supported by a vector potential
\begin{equation}
A_{(1)}={\frac{q}{(d-3)r^{d-3}}}dt,
\end{equation}
with corresponding field strength $F_{(2)}=q\,r^{-(d-2)}\,dt\wedge dr$. An
important difference between the electric and magnetic viewpoints, however,
is that of supersymmetry. While the extremal electrical solutions usually
preserve some fraction of supersymmetry in a corresponding gauged
supergravity theory, this is not generally the case for the magnetic
solutions. Even assuming that the Lagrangian (\ref{eq:mlag}) admits a
supersymmetric generalization, it can be shown that generically the magnetic
black holes break all of the supersymmetries.

This issue with supersymmetry may be illustrated for the special case of
$m=2$, which corresponds to black holes in four dimensions. Here, the metric
function $f(r)$ takes the form
\begin{equation}
f(r)=k-\frac{\omega _{4}M}{r}+\frac{q^{2}}{4r^{2}}+g^{2}r^{2}.
\end{equation}
The 2-form field strength can either be electric, $F_{(2)}=(q/r^{2})\,dt
\wedge dr$, or magnetic, $F_{(2)}=q\,dy^{1}\wedge dy^{2}$. For $k=1$, the
supersymmetric electric $R$-charged black holes may be obtained by taking
the BPS condition $q=\omega _{4}M$, so that $f(r)$ has the extreme
Reissner-Nordstrom-anti de Sitter form
\begin{equation}
f(r)=\left( 1-{\frac{q}{2r}}\right) ^{2}+g^{2}r^{2},\qquad F_{(2)}={\frac{q}
{r^{2}}}dt\wedge dr.
\label{eq:ernads}
\end{equation}
This solution, however, is non-supersymmetric when viewed as a magnetic
black hole. On the other hand, for the particular choice
\begin{equation}
\omega _{4}M=0,\qquad q^{2}={\frac{k^{2}}{g^{2}}},
\label{eq:magquant}
\end{equation}
the magnetic solution has the form
\begin{equation}
f(r)=(gr)^{2}\left( 1+\frac{{k}}{2g^{2}r^{2}}\right) ^{2},\qquad
F_{(2)}=q\,dy^{1}\wedge dy^{2}.
\label{eq:4magbh}
\end{equation}
This is the solution which was found sometime ago by Romans \cite{romans}.
In the context of four-dimensional gauged $N=2$ supergravity, this solution
is BPS and preserves a quarter of the supersymmetry. It was furthermore
shown in \cite{romans} that this `cosmological black hole' solution with
charge quantization given by (\ref{eq:magquant}) is the unique magnetic
solution that preserves some fraction of the supersymmetries.

The extreme $k=1$ Reissner-Nordstrom-anti de Sitter black hole, (\ref
{eq:ernads}), generalizes to arbitrary dimensions. However, the
supersymmetric magnetic solution, (\ref{eq:4magbh}), is restricted to four
dimensions. For five dimensions ($m=3$), there is instead a charged
quantized magnetic black hole solution of the form
\begin{equation}
f(r)=(gr)^{2}\left( 1+{\frac{k}{3g^{2}r^{2}}}\right) ^{3},\qquad
F_{(3)}=q\,dy^{1}\wedge dy^{2}\wedge dy^{3},
\label{eq:5magbh}
\end{equation}
where
\begin{equation}
\omega _{5}M=-\frac{k^{2}}{3g^{2}},\qquad q^{2}=\frac{4k^{3}}{9g^{4}}.
\end{equation}
It would be tempting to view this as a BPS solution to gauged supergravity
in five dimensions. However this interpretation is not at all clear, since
in this case the 3-form field strength ought to satisfy odd-dimensional
self-duality equations \cite
{Townsend:xs,Gunaydin:1984qu,Pernici:ju,Romans:ps}, as opposed to the
standard equations of motion considered here.

It is interesting to note that, for $m\le3$, the charge quantized magnetic
solutions (\ref{eq:4magbh}) and (\ref{eq:5magbh}) follow a harmonic
function-like form
\begin{equation}
f(r)=(gr)^2\left(1+{\frac{k}{m\,g^2r^2}}\right)^m,\qquad m\le3.
\label{eq:genmagbh}
\end{equation}
This no longer holds for $m\ge4$, as then (\ref{eq:ffunc}) can no longer be
factored in this manner.


\section{Magnetically charged Branes}

In this section we are interested in constructing magnetic $p$-brane
solutions for the system described by the Lagrangian (\ref{eq:mlag}). As in
the previous section, it is natural to take a magnetic $m$-form ansatz
$F_{(m)}=q\,d^{m}y$ where $d^{m}y$ is the volume form on an $m$-dimensional
Einstein manifold $\mathcal{M}^{m}$. This motivates us to take a metric
ansatz
\begin{equation}
ds^{2}=e^{2A(r)}g_{\mu \nu }(x)dx^{\mu }dx^{\nu}
+e^{2B(r)}dr^{2}+r^{2}h_{ij}(y)dy^{i}dy^{j}.
\label{eq:bmans}
\end{equation}
Here $\mu ,\nu =1,\ldots ,n$ are $n$ longitudinal directions of a manifold
with metric $g_{\mu \nu }$. The spacetime dimension is thus given by
$d=n+m+1$. For this ansatz, the non-vanishing Ricci components are
\begin{eqnarray}
R_{\mu \nu } &=&R_{\mu \nu }(g)-e^{2A-2B}g_{\mu \nu }\left( A^{\prime \prime
}+nA^{\prime\, 2}-A^{\prime }B^{\prime }+{\frac{m}{r}}A^{\prime }\right) ,
\notag \\
R_{rr} &=&-n\left( A^{\prime \prime }+A^{\prime \,2}-A^{\prime }B^{\prime
}\right) +{\frac{m}{r}}B^{\prime },  \notag \\
R_{ij} &=&R_{ij}(h)+e^{-2B}h_{ij}\left( rB^{\prime }-{nr}A^{\prime }-m+1
\right) .
\end{eqnarray}
This reduces to the previous case, (\ref{eq:n1eins}), when $n=1$. We
now assume the Einstein conditions
\begin{equation}
R_{\mu \nu }(g)=\lambda g_{\mu \nu },\qquad R_{ij}(h)=k(m-1)h_{ij}.
\end{equation}
As a result, for the magnetic $m$-form ansatz, the Einstein equations become
\begin{eqnarray}
A^{\prime \prime }+A^{\prime }(nA^{\prime }-B^{\prime }+{\frac{m}{r}})
&=&e^{2B}\left( {\frac{\left( m-1\right) q^{2}}{2(d-2)r^{2m}}}
+(d-1)g^{2}+\lambda e^{-2A}\right) ,  \notag
\label{eq:eom2} \\
n(A^{\prime \prime }+A^{\prime }{}^{2}-A^{\prime }B^{\prime })-{\frac{m}{r}}
B^{\prime } &=&e^{2B}\left( {\frac{\left( m-1\right) q^{2}}{2(d-2)r^{2m}}}
+(d-1)g^{2}\right) , \\
{\frac{1}{r}}(nA^{\prime }-B^{\prime })+{\frac{m-1}{r^{2}}} &=&e^{2B}\left(
-{\frac{nq^{2}}{2(d-2)r^{2m}}}+(d-1)g^{2}+{\frac{k(m-1)}{r^{2}}}\right) .
\notag
\end{eqnarray}

Before proceeding, we note that these expressions may be slightly
simplified. Subtracting the first two equations gives
\begin{equation}
A^{\prime \prime }-A^{\prime }B^{\prime }-{\frac{m}{\left( n-1\right) r}}
(A^{\prime }+B^{\prime })=-{\frac{\lambda }{n-1}}e^{2B-2A}.
\label{eq:abeqn}
\end{equation}
Alternatively, we may eliminate $A^{\prime \prime }$ from the first two
equations to obtain
\begin{equation}
nA^{\prime }{}^{2}+{\frac{m}{\left( n-1\right) r}}(nA^{\prime }+B^{\prime })=
{\frac{n\lambda }{n-1}}e^{2B-2A}+e^{2B}\left( {\frac{\left( m-1\right) q^{2}}
{2(d-2)r^{2m}}}+(d-1)g^{2}\right).
\label{eq:e1asq}
\end{equation}
We are thus left with two coupled non-linear first order equations, namely
(\ref{eq:e1asq}) and the last equation of (\ref{eq:eom2}), as well as a
second order equation, (\ref{eq:abeqn}). However, the second order equation
is in fact redundant, as it may be obtained by differentiation of the first
order equations. To see this explicitly, we may first eliminate $B^{\prime }$
from the first order equations to obtain
\begin{eqnarray}
&&e^{-2B}\left( n(n-1)r^{2}A^{\prime \,2}+2mnrA^{\prime }+m(m-1)\right) =
\notag \\
&&\kern8emn\lambda r^{2}e^{-2A}-{\frac{1}{2}}{\frac{q^{2}}{r^{2(m-1)}}}
+(d-1)(d-2)g^{2}r^{2}+km(m-1).\quad
\end{eqnarray}
Taking a derivative and using (\ref{eq:e1asq}) to eliminate the term
proportional to $q^{2}$, we find
\begin{equation}
2n\left( m+(n-1)rA^{\prime }\right) \left( A^{\prime \prime }-A^{\prime
}B^{\prime }-{\frac{m}{n-1}}{\frac{1}{r}}(A^{\prime }+B^{\prime })
+{\frac{\lambda }{n-1}}e^{2B-2A}\right) =0,
\end{equation}
which proves the claim, at least provided $m+(n-1)rA^{\prime }\neq 0$, or
equivalently $e^{2A}\neq c_{0}r^{-(2m)/(n-1)}$ (for some constant $c_{0}$).

Our goal, thus, is to solve the coupled first order equations for $A(r)$ and
$B(r)$ with appropriate boundary conditions. However, before we do so, let
us note that for vanishing magnetic charge, the Einstein equation is simply
$R_{MN}=-(d-1)g^{2}g_{MN}$. However, the symmetry of the ansatz, (\ref
{eq:bmans}), precludes the maximally symmetric AdS$_{d}$ vacuum. Instead, we
note that the Einstein condition may be solved by choosing $\lambda
=(m-1)g^{2}k$, so that the metric (for $q=0$) takes on the form
\begin{equation}
ds^{2}={\frac{dr^{2}}{\hat{k}+g^{2}r^{2}}}+r^{2}\left( g^{2}g_{\mu \nu
}(x)dx^{\mu }dx^{\nu }+h_{ij}(y)dy^{i}dy^{j}\right) ,
\end{equation}
where $\hat{k}=(m-1)k/(d-2)$.

Although this vacuum would not be supersymmetric (in a gauged supergravity
context), it nevertheless leads to an asymptotically anti-de Sitter geometry
of the form
\begin{equation}
ds^2\approx g^2r^2(g_{\mu\nu}dx^\mu dx^\nu+g^{-2}h_{ij}dy^idy^j) +g^{-2}
{\frac{dr^2}{r^2}}.
\label{eq:asympmet}
\end{equation}
For non-zero magnetic charge, we note that, asymptotically the magnetic
field strength, (\ref{eq:fsqs}), falls off as $F_\m^2\sim 1/r^{2m}$. Thus
even when $q\ne0$, it is natural to expect the magnetic brane solution to be
asymptotic to (\ref{eq:asympmet}), at least up to corrections of the order
$\mathcal{O}(1/r^2)$. Hence we require that
\begin{equation}
e^{2A}\sim e^{-2B}\sim (gr)^2\qquad\hbox{as}\qquad r\to\infty.
\label{eq:asympbc}
\end{equation}

Formally, the system of first order equations, (\ref{eq:e1asq}) and the last
equation of (\ref{eq:eom2}), admits a two-parameter family of solutions; in
addition, there are of course the inputs $\lambda$ (the cosmological
constant on the longitudinal space) and $q$ (the magnetic charge) as well as
the discrete parameter $k=1,0,-1$.
While we have been unable to find an explicit solution for arbitrary values
of $\lambda$ and $q$, we note that the equations simplify if we take the
longitudinal space to the Ricci-flat, $\lambda=0$. In this case, we may
combine (\ref{eq:e1asq}) with the last equation of (\ref{eq:eom2}) to
eliminate $A^{\prime}$. This results in a first order equation
\begin{eqnarray}
&&\kern-10pt f^{\prime\,2}-{\frac{4m}{n-1}}{\frac{1}{r}}ff^{\prime}
-{\frac{4m(d-1)}{n-1}}{\frac{1}{r^{2}}}f^{2} -{\frac{4}{r}}f^{\prime}
\left((d-1)+{\frac{k(m-1)}{(gr)^2}} -{\frac{\widetilde q\,^2}{(gr)^{2m}}}
\right)  \notag \\
&&\kern-10pt +{\frac{4}{n-1}}{\frac{1}{r^{2}}}f \left((d-1)(m-(d-1)(n-1))
-{\frac{km(m-1)(n-2)}{(gr)^2}} +(n-m-1){\frac{\widetilde q\,^2}{(gr)^{2m}}}
\right)  \notag \\
&&\kern-10pt +{\frac{4}{r^{2}}} \left((d-1)+{\frac{k(m-1)}{(gr)^2}}
-{\frac{\widetilde q\,^2}{(gr)^{2m}}}\right)^2=0,
\label{eq:diffeq}
\end{eqnarray}
where
\begin{equation}
f={\frac{1}{(gr)^2}}e^{-2B},
\label{eq:fdef}
\end{equation}
and $\widetilde q$ is a rescaled (dimensionless) magnetic charge
\begin{equation}
\widetilde q\,^2={\frac{n}{2(d-2)}}g^{2(m-1)}q^2.
\end{equation}
In general one may seek numerical solutions to (\ref{eq:diffeq}) subject to
the boundary condition $f(r)\to1$ as $r\to\infty$. However closed form
solutions may be obtained for the special cases of $m=2$ and $m=3$, provided
a form of charge quantization is imposed.

The $m=2$ ($d=n+3$) case, corresponding to Einstein-Maxwell theory with a
cosmological constant, has recently been investigated in \cite
{sabra2,sabra2a}. Given the charge quantization condition
\begin{equation}
\widetilde{q}\,^{2}={\frac{k^{2}}{(n+1)^{2}}},
\end{equation}
we find a solution to (\ref{eq:diffeq}) of the form
\begin{equation}
e^{-2B}=(gr)^{2}\left(1+{\frac{k}{(n+1)g^{2}r^{2}}}\right)^{2}.
\label{eq:m2sol1}
\end{equation}
This may then be inserted into the last equation of (\ref{eq:eom2}) to
obtain an equation for $A$, which may be solved to give
\begin{equation}
e^{2A}=(gr)^{2}\left( 1+{\frac{k}{(n+1)g^{2}r^{2}}} \right)^{{\frac{n+1}{n}}}.
\label{eq:m2sol2}
\end{equation}
This agrees with the magnetic co-dimension three solutions obtained in \cite
{sabra2,sabra2a}.

Turning now to the $m=3$ case, we find that a closed form solution exists
provided we impose a charge quantization condition
\begin{equation}
\widetilde{q}\,^{2}={\frac{16n^{2}k^{3}}{(n+1)^{3}(n+2)^{3}}}.
\label{eq:m3quant}
\end{equation}
This time, the solution for $B$ is somewhat more complicated, and is given
by
\begin{equation}
e^{-2B}=(gr)^{2}\left( 1+{\frac{2k}{(n+1)(n+2)g^{2}r^{2}}}\right) \left(1
+{\frac{2nk}{(n+1)(n+2)g^{2}r^{2}}}\right)^2,
\label{eq:m3bsoln}
\end{equation}
whereas that for $A$ has the simple form
\begin{equation}
e^{2A}=(gr)^{2}\left( 1+\frac{2nk}{(n+1)(n+2)g^{2}r^{2}}\right)
^{\frac{n+2}{n}}.
\label{eq:m3asoln}
\end{equation}
Note that for $n=1$, both the $m=2$ and $m=3$ solutions reduce to magnetic
black holes with quantized magnetic charge, given by (\ref{eq:genmagbh}).
Thus these solutions may be viewed as a larger family of `cosmological
magnetic branes' generalizing the cosmological black holes of \cite{romans}.

Although we have searched for a generalization to arbitrary $m$, we have as
yet been unsuccessful. In particular, it can be shown that no terminating
series solution for $f$ given in (\ref{eq:fdef}) exists for $m\geq 4$,
regardless of the charge $\widetilde q$, while assuming the asymptotics of
(\ref{eq:asympbc}). Of course, this result is not entirely unexpected, as
even for $n=1$, factorization of the metric function, (\ref{eq:genmagbh}),
is only possible for $m\le3$.


\section{The near-horizon solution}

In the previous sections, we have considered a transverse space metric of
the form
\begin{equation}
ds^{2}=e^{2B}dr^{2}+r^{2}d\Omega_k^2,
\end{equation}
corresponding to a foliation of space in terms of hypersurfaces with
curvature $k=1$, $0$ or $-1$ (corresponding to the horizon geometry of the
black object). On the other hand, assuming a regular horizon with finite
area (\textit{i.e.} horizon radius $r_{0}>0$), it is instructive to examine
the near horizon solution. In this case, it is natural to take a direct
product metric
\begin{equation}
ds^{2}=g_{\mu \nu }(x)dx^{\mu }dx^{\nu }+h_{ij}(y)dy^{i}dy^{j},
\end{equation}
where $\mu ,\nu =1,\ldots ,d-m$ and $i,j=1,\ldots ,m$. This essentially
corresponds to a Freund-Rubin compactification with magnetic field strength
given by
\begin{equation}
F_{(m)}=q\,d^{m}y.
\end{equation}
Similar to (\ref{eq:fsqs}), this yields
\begin{equation}
F_{(m)\,ij}^{2}=(m-1)!\,q^{2}h_{ij},\qquad F_{(m)}^{2}=m!\,q^{2}.
\end{equation}
The resulting Einstein equations are then
\begin{eqnarray}
R_{\mu \nu }(x) &=&-\left[ {\frac{m-1}{2(d-2)}}q^{2}+(d-1)g^{2}\right]
g_{\mu\nu}, \notag \\
R_{ij}(y) &=&\left[ {\frac{d-m-1}{2(d-2)}}q^{2}-(d-1)g^{2}\right] h_{ij}.
\label{eq:freom}
\end{eqnarray}
As a consequence, the scalar curvature satisfies
\begin{equation}
R={\frac{d-2m}{2(d-2)}}q^{2}-d(d-1)g^{2}.
\end{equation}

The equations, (\ref{eq:freom}), are simply Einstein conditions for the
`longitudinal' and `transverse' dimensions, with cosmological constants
\begin{eqnarray}
\Lambda _{d-m} &=&-\left[ {\frac{m-1}{2(d-2)}}q^{2}+(d-1)g^{2}\right],
\notag\\
\Lambda _{m} &=&\left[ {\frac{d-m-1}{2(d-2)}}q^{2}-(d-1)g^{2}\right] .
\end{eqnarray}
Note that the longitudinal space is always AdS (since $\Lambda _{d-m}<0$).
However, the transverse space may have either sign for $\Lambda _{m}$,
depending on the relative strength of $q^{2}$ versus $g^{2}$. It is also
possible to set $\Lambda _{m}$ to zero, although without supersymmetry this
may only be viewed as a fine tuning. The freedom to adjust $\Lambda _{m}$
may be understood by recalling that, starting from a $d$-dimensional theory
without cosmological constant, the Freund-Rubin compactification yields a
geometry of the form AdS $\times $ Sphere. Turning on an overall negative
cosmological constant in the initial higher dimensional theory then
contributes an additional negative factor for both the AdS and sphere
curvatures. For a sufficiently large negative cosmological constant, the
sphere is then replaced by hyperbolic space.

Of course, since we are considering non-supersymmetric models, we could
equally well have started with a positive cosmological constant in the
$d$-dimensional theory. In this case, turning on fluxes would yield a
longitudinal space of arbitrary curvature, while the transverse space is
always positively curved. In this case, an appropriate fine tuning would
yield a geometry of the form Minkowski $\times $ Sphere.


\section{Cosmological solutions}

Until now, we have been considering only static magnetic AdS brane
solutions. However, it is straightforward to analyze time dependent de
Sitter solutions as well. Here we will obtain a set of cosmological
solutions to the system described by the Lagrangian
\begin{equation}
e^{-1}\mathcal{L}=R-{\frac{1}{2\cdot m!}}F_{(m)}^{2}-(d-1)(d-2)g^{2}.
\end{equation}
This is identical to (\ref{eq:mlag}), except that here the cosmological
constant is positive. As a metric
ansatz for the time-dependent solutions, we take our total spacetime
dimension to be $d=(n+1)+m$ and assume a factorized form $K^{1,n}\times
\mathcal{M}^{m}$ where $K^{1,n}$ is spatially isotropic and $\mathcal{M}^{m}$
is Einstein, with metric $h_{ij}$ with zero, positive or negative curvature.
To be precise, the ansatz is given by
\begin{equation}
ds^{2}=-e^{2B(t)}dt^{2}+e^{2A(t)}\left( dx_{1}^{2}+\cdots +dx_{n}^{2}\right)
+t^{2}h_{ij}dy^{i}dy^{j},
\label{eq:cosans}
\end{equation}
where $i,j=1,\ldots ,m$. This is a natural cosmological version of the
original static ansatz given by (\ref{eq:bmans}). For a magnetic solution,
we again take $F_{(m)}=q\,d^{m}y$. For this ansatz, the non-vanishing Ricci
components are
\begin{eqnarray}
R_{tt} &=&-n(\ddot{A}+\dot{A}^{2}-\dot{A}\dot{B})+\frac{m}{t}\dot{B},
\notag\\
R_{ab} &=&e^{2A-2B}\delta _{ab}\left( \ddot{A}+n\dot{A}^{2}-\dot{A}
\dot{B}+\frac{m}{t}\dot{A}\right) ,  \notag \\
R_{ij} &=&h_{ij}\left[e^{-2B}\left( m-1-t\dot{B}+nt\dot{A}\right)
+k(m-1)\right],
\label{Rabt}
\end{eqnarray}
so that the Einstein equations take the form
\begin{eqnarray}
-n(\ddot{A}+\dot{A}^{2}-\dot{A}\dot{B})+\frac{m}{t}\dot{B} &=&e^{2B}\left(
{\frac{\left( m-1\right) }{2(d-2)}\frac{q^{2}}{t^{2m}}}-(d-1)g^{2}\right),
\notag \\
-\left( \ddot{A}+n\dot{A}^{2}-\dot{A}\dot{B}+\frac{m}{t}\dot{A}\right)
&=&e^{2B}\left( {\frac{\left( m-1\right) }{2(d-2)}\frac{q^{2}}{t^{2m}}}
-(d-1)g^{2}\right), \notag \\
\frac{1}{t}(n\dot{A}-\dot{B})+\frac{m-1}{t^{2}} &=&e^{2B}
\left({\frac{n}{2(d-2)}\frac{q^{2}}{t^{2m}}}+(d-1)g^{2}
-\frac{k(m-1)}{t^{2}}\right).
\label{e1}
\end{eqnarray}

For the special case $n=1$, the first two equations imply $\dot{A}=-\dot{B}$
and the solution is given by
\begin{equation}
ds^{2}=-{\frac{dt^{2}}{f(t)}}+f(t)dr^{2}+t^{2}h_{ij}(y)dy^{i}dy^{j},
\label{eq:dsbh1}
\end{equation}
where
\begin{equation}
f(t)=-k-\frac{\mu }{t^{d-3}}-\frac{q^{2}}{2(d-2)(d-3)}\frac{1}{t^{2(d-3)}}
+g^{2}t^{2}.
\label{eq:dsbh2}
\end{equation}
For the choice $k=1$ and $q^{2}=\frac{\mu ^{2}}{2}\left( d-2\right) \left(
d-3\right) $, the function $f(t)$ becomes
\begin{equation}
f(t)=g^{2}t^{2}-\left( 1-{\frac{\mu }{t^{d-3}}}\right) ^{2},
\end{equation}
and in this case one obtains the solution presented in \cite{hongt}.

Of course, it should be noted that the cosmological solution (\ref{eq:dsbh1})
and (\ref{eq:dsbh2}), is formally identical to that of the static
Reissner-Nordstrom-de Sitter black hole
\begin{eqnarray}
&&ds^2=-g(r)dt^2+{\frac{dr^2}{g(r)}}+r^2h_{ij}dy^idy^j,  \notag \\
&&g(r)=k-{\frac{\mu}{r^{d-3}}}+{\frac{q^2}{2(d-2)(d-3)}}{\frac{1}{r^{2(d-3)}}}
-g^2r^2.
\end{eqnarray}
This may be seen by making the substitution $r\leftrightarrow t$ as well as
$f(t)\to -g(r)$ and $\mu\to-\mu$. However, here we are interested in the
region of spacetime given by $f(t)>0$. This corresponds to the counterpart
of the static black hole solution on the other side of the de Sitter
horizon. In other words, for the time dependent solution, we focus on the
region of spacetime given by $g(r)<0$, where the r\^oles of $r$ and $t$ are
interchanged.

For $m=2$, namely Einstein-Maxwell-de Sitter, we get the following solution
\cite{hongds1}
\begin{equation}
ds^{2}=-(gt)^{-2}\left( 1-{\ \frac{k}{(n+1)g^{2}t^{2}}}\right)
^{-2}dt^{2}+(gt)^{2}\left( 1-{\frac{k}{(n+1)g^{2}t^{2}}}\right)
^{{\frac{n+1}{n}}}d\vec{x}\,^{2}+t^{2}d\Omega _{k}^{2},
\end{equation}
where $d\Omega _{k}^{2}$ is the metric of a two-dimensional manifold with
$k=1,0,-1$. However, the charge quantization is given by
\begin{equation}
q^{2}=-{\frac{2k^{2}}{g^{2}n(n+1)}}.
\end{equation}
Thus this solution has imaginary magnetic charge, and for $n=2$ may be
viewed as a solution of the five dimensional supergravity theory arising
from the reduction of IIB* on $dS_{5}\times H^{5}$. These reductions yield
five-dimensional de Sitter supergravities, albeit with wrong sign kinetic
terms \cite{jww}.

For the case $m=3$, we obtain a time-dependent solution in de Sitter space
\begin{eqnarray}
ds^{2} &=&-(gt)^{-2}\left( 1-{\frac{2k}{(n+1)(n+2)g^{2}t^{2}}}\right)
^{-1}\left( 1-{\frac{2nk}{(n+1)(n+2)g^{2}t^{2}}}\right) ^{-2}dt^{2}
\notag\\
&&+(gt)^{2}\left( 1-{\frac{2nk}{(n+1)(n+2)g^{2}t^{2}}}\right)
^{{\ \frac{n+2}{n}}}d\vec{x}\,^{2}+t^{2}h_{ij}(y)dy^{i}dy^{j},
\end{eqnarray}
with magnetic flux satisfying
\begin{equation}
q^{2}=\frac{32nk^{3}}{g^{4}(n+1)^{3}(n+2)^{2}}.
\end{equation}
This charge quantization condition is identical in sign with that for the
static anti-de Sitter case, namely (\ref{eq:m3quant}). Thus one must take
$k=1$ in order to obtain real gauge fields in both the AdS and dS cases.

\section{Discussion}

In this paper, we have focused on constructing magnetic brane solutions
asymptotic to anti-de Sitter space. For the $0$-brane case, we recover the
magnetically charged Reissner-Nordstrom-AdS black hole. On the other hand,
we have demonstrated that a new class of magnetic $p$-branes ($p>0$) may be
obtained by solution of the non-linear equation (\ref{eq:fdef}). For the
case of $m=2$ (Einstein-Maxwell), the co-dimension three branes reproduce
the solutions of \cite{sabra2,sabra2a}, while for $m=3$, we obtain a new
analytic solution for co-dimension four branes.

In general, we are unable to work in a supersymmetric framework. This is due
to the fact that gauged supergravities (which yield AdS vacua) generally do
not contain $m$-form field strengths with $m>2$ satisfying standard $m$-form
equations of motion. Instead, such higher form gauge fields satisfy
odd-dimensional self-duality equations, at least in the five and
seven-dimensional gauged supergravities \cite
{Townsend:xs,Gunaydin:1984qu,Pernici:ju,Romans:ps}. On the other hand, for
$m=2$, the Lagrangian (\ref{eq:mlag}) may generally be given a supersymmetric
interpretation with $F_{(2)}$ transforming as a graviphoton. In this case,
it is known that the extreme electrically charged AdS black holes preserve a
fraction of the supersymmetries \cite
{romans,blackholes,bha,bhb,bhc,Liu:1999ai,bhi}. For the magnetic objects,
the charge quantized co-dimension three solution (\ref{eq:m2sol1}) and (\ref
{eq:m2sol2}) was shown to be one quarter supersymmetric, both in four
dimensions \cite{romans} and in general \cite{sabra2a}.

Although we are not aware of a gauged supergravity with $m=3$ and a standard
kinetic term of the form (\ref{eq:mlag}), it is tempting to suggest that the
cosmological magnetic brane solution (\ref{eq:m3bsoln}) and (\ref{eq:m3asoln})
is likewise somehow one quarter supersymmetric. In five dimensions, one
may be tempted to identify this magnetic $F_{(3)}$ black hole as dual to an
electric $F_{(2)}$ black hole (which would be supersymmetric in the extremal
limit). However, this cannot be the case, as it is not possible to dualize
the graviphotons in gauged supergravity. Nevertheless, we hold open the
possibility that some dual five-dimensional theory may be formulated where
the gauging is accomplished via an antisymmetric tensor with field strength
$F_{(3)}=dA_{(2)}$.

Finally, we note that a class of supersymmetric branes supported by higher
form potentials have been constructed using the L\"u-Pope ansatz \cite
{LP,CLP,FP,steve}. These solutions make explicit use of the odd-dimensional
self-duality equations. Thus they appear rather different in structure from
the magnetic branes considered above. Furthermore, the L\"u-Pope lifted
solutions have both electric and magnetic components of the field strengths
active by virtue of odd-dimensional self-duality. This suggests that, in
order to make a connection between the solutions discussed here and the
branes of the L\"u-Pope type, we would at least have to generalize our
magnetic ansatz to include the dyonic case. It would be interesting to see
if this is indeed possible.

\section*{Acknowledgments}

This work was supported in part by the US Department of Energy under grant
DE-FG02-95ER40899. JTL acknowledges the hospitality of Khuri lab at the
Rockefeller University where this project was initiated and the theory group
at Princeton University, where part of this work was completed.


\end{document}